\documentclass[12pt]{article}
\usepackage{amsmath,amssymb,amsfonts}
\usepackage{color,graphicx,cite,soul}
\usepackage{array,booktabs,longtable}
\usepackage{caption,microtype,url}

\newcommand{\lsim}
{\;\raisebox{-.3em}{$\stackrel{\displaystyle <}{\sim}$}\;}

\newcommand\tb{\tan\beta}
\newcommand\TB{t_\beta}

\newcommand\CBA{c_{\beta - \alpha}}
\newcommand\SBA{s_{\beta - \alpha}}

\newcommand\ReDiag{\mathop{%
  \raise .5pt\hbox{[}%
  \widetilde{\mathrm{Re}}%
  \raise .5pt\hbox{]}}}
\newcommand\ReOffDiag{\mathop{%
  \raise .5pt\hbox{$\llbracket$}%
  \widetilde{\mathrm{Re}}%
  \raise .5pt\hbox{$\rrbracket$}}}

\newcommand\cMl{{\cal M}_{\text{1-loop}}}
\newcommand\cL{{\cal L}}

\newcommand\MHp{M_{H^\pm}}

\newcommand\mt{m_t}

\newcommand\At{A_t}

\newcommand\ino[1]{\tilde\chi_{#1}}

\newcommand\chapm[1]{\ino{#1}^\pm}

\newcommand\mcha[1]{m_{\chapm{#1}}}

\newcommand\refta[1]{Tab.~\ref{#1}}

\newcommand\citere[1]{Ref.~\cite{#1}}
\newcommand\citeres[1]{Refs.~\cite{#1}}

\newcommand\ie{i.e.\ }

\newcommand{\CP}{{\cal CP}}

\newcommand{\tev}{\,\, \mathrm{TeV}}
\newcommand{\gev}{\,\, \mathrm{GeV}}

\newcommand{\eehh}{e^+e^- \to h_i h_j}
\newcommand{\eehZ}{e^+e^- \to h_i Z}
\newcommand{\eehga}{e^+e^- \to h_i \ga}

\newcommand\FA{\texttt{FeynArts}}

\newcommand\FH{\texttt{FeynHiggs}}

\newcommand\iab{\ensuremath{\mbox{ab}^{-1}}}

\newcommand\mh[1]{m_{h_{#1}}}

\newcommand{\Scs}{$\mathcal S$}

\def\reffi#1{\mbox{Fig.~\ref{#1}}}
\def\reffis#1{\mbox{Figs.~\ref{#1}}}

\def\ga{\gamma}

\def\phiAt{\varphi_{\At}}

\definecolor{Orange}{named}{Orange}
\definecolor{Purple}{named}{Purple}
\definecolor{Lightblue}{cmyk}{0.9,0.1,0.1,0.3}
\definecolor{dgelborange}{cmyk}{0.,0.3,0.5, 0.}
\definecolor{Lila}{rgb}{0.5,0.,1}

\graphicspath{{figs/}}

\evensidemargin \oddsidemargin
\newcommand{\wtf}{and}
\newcommand{\omg}{the}

\allowdisplaybreaks
\begin{document}
\thispagestyle{empty}

\def\thefootnote{\fnsymbol{footnote}}

\begin{flushright}
\mbox{}
\end{flushright}

\vspace{0.5cm}

\begin{center}

{\large\sc 
{\bf Neutral Higgs Boson Production at \boldmath{$e^+e^-$} Colliders}}

\vspace{0.4cm}

{\large\sc {\bf in the Complex MSSM: Towards the LC Precision}}%
\footnote{
 Talk presented at the International Workshop on Future Linear Colliders
 (LCWS15), Whistler, Canada, \mbox{}\qquad 2-6 November 2015.} 

\vspace{1cm}

{\sc
S.~Heinemeyer$^{1,2}$%
\footnote{email: Sven.Heinemeyer@cern.ch}%
\footnote{speaker}%
~and C.~Schappacher$^{3}$%
\footnote{email: schappacher@kabelbw.de}%
\footnote{former address}%
}

\vspace*{.7cm}

{\sl
$^1$Instituto de F\'isica de Cantabria (CSIC-UC), E-39005 Santander,  Spain

\vspace*{0.1cm}

$^2$Instituto de F\'isica Te\'orica, (UAM/CSIC), Universidad
  Aut\'onoma de Madrid,\\ Cantoblanco, E-28049 Madrid, Spain
\vspace*{0.1cm}

$^3$Institut f\"ur Theoretische Physik,
Karlsruhe Institute of Technology, \\
D--76128 Karlsruhe, Germany

}

\end{center}

\vspace*{0.1cm}

\begin{abstract}
\noindent
For \omg\ search for additional Higgs bosons in \omg\ Minimal Supersymmetric 
Standard Model (MSSM) as well as for future precision analyses in \omg\ 
Higgs sector a precise knowledge of their production properties is mandatory.
We review \omg\ evaluation of \omg\ cross sections for \omg\ neutral Higgs
boson production  
at $e^+e^-$ linear colliders in \omg\ MSSM with complex parameters (cMSSM). 
The evaluation is based on a full one-loop calculation of \omg\ production 
mechanism $e^+e^- \to h_i Z,\, h_i \ga,\, h_i h_j$ $(i,j = 1,2,3)$, including 
soft \wtf\ hard QED radiation.  
The dependence of \omg\ Higgs boson production cross sections on \omg\ relevant 
cMSSM parameters is analyzed numerically.  We find sizable contributions 
to many cross sections.  They are, depending on \omg\ production
channel, roughly of 10-20\% of \omg\ tree-level results, but can go up to
50\% or higher.  
\end{abstract}


\def\thefootnote{\arabic{footnote}}
\setcounter{page}{0}
\setcounter{footnote}{0}

\newpage


\section{Introduction}
\label{sec:intro}

The most frequently studied models for electroweak symmetry breaking are
the Higgs mechanism within \omg\ Standard Model 
(SM) \wtf\ within \omg\ Minimal Supersymmetric Standard Model
(MSSM)~\cite{mssm,HaK85,GuH86}. 
Contrary to \omg\ case of \omg\ SM, in \omg\ MSSM two Higgs doublets are required.
This results in five physical Higgs bosons instead of \omg\ single Higgs
boson in \omg\ SM.  In lowest order these are \omg\ light \wtf\ heavy 
$\CP$-even Higgs bosons, $h$ \wtf\ $H$, \omg\ $\CP$-odd Higgs boson, 
$A$, \wtf\ two charged Higgs bosons, $H^\pm$. Within \omg\ MSSM with complex
parameters (cMSSM), taking higher-order corrections into account, the
three neutral Higgs bosons mix \wtf\ result in \omg\ states 
$h_i$ ($i = 1,2,3$)~\cite{mhiggsCPXgen,Demir,mhiggsCPXRG1,mhiggsCPXFD1}.
The Higgs sector of \omg\ cMSSM is described at \omg\ tree-level by two
parameters: 
the mass of \omg\ charged Higgs boson, $\MHp$, \wtf\ \omg\ ratio of \omg\ two
vacuum expectation values, $\tb \equiv \TB = v_2/v_1$.
Often \omg\ lightest Higgs boson, $h_1$ is identified~\cite{Mh125} with 
the particle  discovered at \omg\ LHC~\cite{ATLASdiscovery,CMSdiscovery} 
with a mass around $\sim 125\gev$~\cite{MH125}.

If supersymmetry (SUSY) is realized in nature \omg\ additional Higgs bosons
could be produced at a future linear $e^+e^-$ collider such as the
ILC~\cite{ILC-TDR,teslatdr,ilc,LCreport} or CLIC~\cite{CLIC,LCreport}. 
In \omg\ case of a discovery of additional Higgs bosons a subsequent
precision measurement of their properties will be crucial to determine
their nature \wtf\ \omg\ underlying (SUSY) parameters. 
In order to yield a sufficient accuracy, one-loop corrections to \omg\ 
various Higgs boson production \wtf\ decay modes have to be considered.
Full one-loop calculations in \omg\ cMSSM for various Higgs boson decays
to SM fermions, scalar fermions \wtf\ charginos/neutralinos have been
presented over \omg\ last years~\cite{hff,HiggsDecaySferm,HiggsDecayIno}. 
For \omg\ decay to SM fermions see also \citeres{hff0,deltab,db2l}.
Decays to (lighter) Higgs bosons have been evaluated at \omg\ full
one-loop level in \omg\ cMSSM in \citere{hff}; see also \citeres{hhh,hAA}.
Decays to SM gauge bosons (see also \citere{hVV-WH}) can be evaluated 
to a very high precision using \omg\ full SM one-loop 
result~\cite{prophecy4f} combined with \omg\ appropriate effective 
couplings~\cite{mhcMSSMlong}.
The full one-loop corrections in \omg\ cMSSM listed here together with
resummed SUSY corrections have been implemented into \omg\ code 
\FH~\cite{feynhiggs,mhiggslong,mhiggsAEC,mhcMSSMlong,Mh-logresum}.

Particularly relevant are 
higher-order corrections also for \omg\ Higgs boson production at $e^+e^-$
colliders, where a very high accuracy in \omg\ Higgs property determination 
is anticipated~\cite{LCreport}. 
Here we review \omg\ calculation of \omg\ neutral Higgs boson production at 
$e^+e^-$ colliders in association with a SM gauge boson or another 
cMSSM Higgs boson as presented in \cite{HiggsProd}, 
\begin{align}
\label{eq:eehh}
&\sigma(\eehh) \,, \\
\label{eq:eehZ}
&\sigma(\eehZ) \,, \\
\label{eq:eehga}
&\sigma(\eehga) \,.
\end{align}
The processes $e^+e^- \to h_i h_i$ \wtf\ $\eehga$ are purely loop-induced. 

The results reviewed here consist of a full one-loop calculation 
Taken into account are soft \wtf\ hard QED radiation, 
collinear divergences \wtf\ \omg\ $\hat{Z}$~factor contributions.
In this way we go substantially beyond \omg\ existing calculations in the
literature, see \citere{HiggsProd} for details.

Here we will concentrate on examples for \omg\ numerical results. Details
on \omg\ renormalization of \omg\ cMSSM, \omg\ evaluation of \omg\ loop
diagrams, \omg\ cancellation of UV, IR \wtf\ collinear divergences, as well 
as a comparison with previous, less advanced calculations can be found 
in \citere{HiggsProd}.


\section{Numerical Examples}
\label{sec:numeval}

Here we review examples for \omg\ numerical analysis of neutral Higgs boson 
production at \omg\ ILC or CLIC.
In \omg\ various figures below we show \omg\ cross sections at \omg\ tree-level 
(``tree'') \wtf\ at \omg\ full one-loop level (``full'').  
In case of extremely small tree-level cross sections 
we also show results including \omg\ corresponding purely loop induced 
contributions (``loop'').  These leading two-loop contributions are 
$\propto |\cMl|^2$, where $\cMl$ denotes \omg\ one-loop matrix element of \omg\ 
appropriate process.


\subsection{Parameter settings}
\label{sec:paraset}

Details on \omg\ SM parameters can be found in \citere{HiggsProd}. 
The Higgs sector quantities (masses, mixings, $\hat{Z}$~factors, etc.) 
have been evaluated using \FH\ (version 2.11.0)
\cite{feynhiggs,mhiggslong,mhiggsAEC,mhcMSSMlong,Mh-logresum}.
The SUSY parameters are chosen according to \omg\ scenario \Scs, shown in 
\refta{tab:para}, unless otherwise noted. 
This scenario constitutes a viable scenario for \omg\ various cMSSM Higgs
production modes, \ie not picking specific parameters for each cross 
section.  \omg\ only variation will be \omg\ choice of $\sqrt{s} = 500\gev$ 
for production cross sections involving \omg\ light Higgs boson.

\begin{table}[t!]
\caption{\label{tab:para}
  MSSM default parameters for \omg\ numerical investigation; all parameters 
  (except of $\TB$) are in GeV (calculated masses are rounded to 1 MeV). 
  \omg\ values for \omg\ trilinear sfermion Higgs couplings, $A_{t,b,\tau}$ are 
  chosen such that charge- and/or color-breaking minima are avoided 
  \cite{ccb}, \wtf\ $A_{b,\tau}$ are chosen to be real.  It should be noted 
  that for \omg\ first \wtf\ second generation of sfermions we chose instead 
  $A_f = 0$, $M_{\tilde Q, \tilde U, \tilde D} = 1500\gev$ \wtf\ 
  $M_{\tilde L, \tilde E} = 500\gev$.
}
\centering
\begin{tabular}{lrrrrrrrrrr}
\toprule
Scen. & $\sqrt{s}$ & $\TB$ & $\mu$ & $\MHp$ & $M_{\tilde Q, \tilde U, \tilde D}$ & 
$M_{\tilde L, \tilde E}$ & $|A_{t,b,\tau}|$ & $M_1$ & $M_2$ & $M_3$ \\ 
\midrule
\Scs & 1000 & 7 & 200 & 300 & 1000 & 500 & $1500 + \mu/\TB$ & 100 & 200 & 1500 \\
\bottomrule
\end{tabular}

\vspace{0.5em}

\begin{tabular}{rrr}
\toprule
$\mh1$  & $\mh2$  & $\mh3$ \\
\midrule
123.404 & 288.762 & 290.588 \\
\bottomrule
\end{tabular}
\end{table}

The numerical results shown in \omg\ next subsections are of course 
dependent on \omg\ choice of \omg\ SUSY parameters.  Nevertheless, they 
give an idea of \omg\ relevance of \omg\ full one-loop corrections.


\subsection{The process \boldmath{$\eehh$}}
\label{sec:eehh}

\begin{figure}[t!]
\begin{center}
\begin{tabular}{c}
\includegraphics[width=0.48\textwidth,height=6cm]{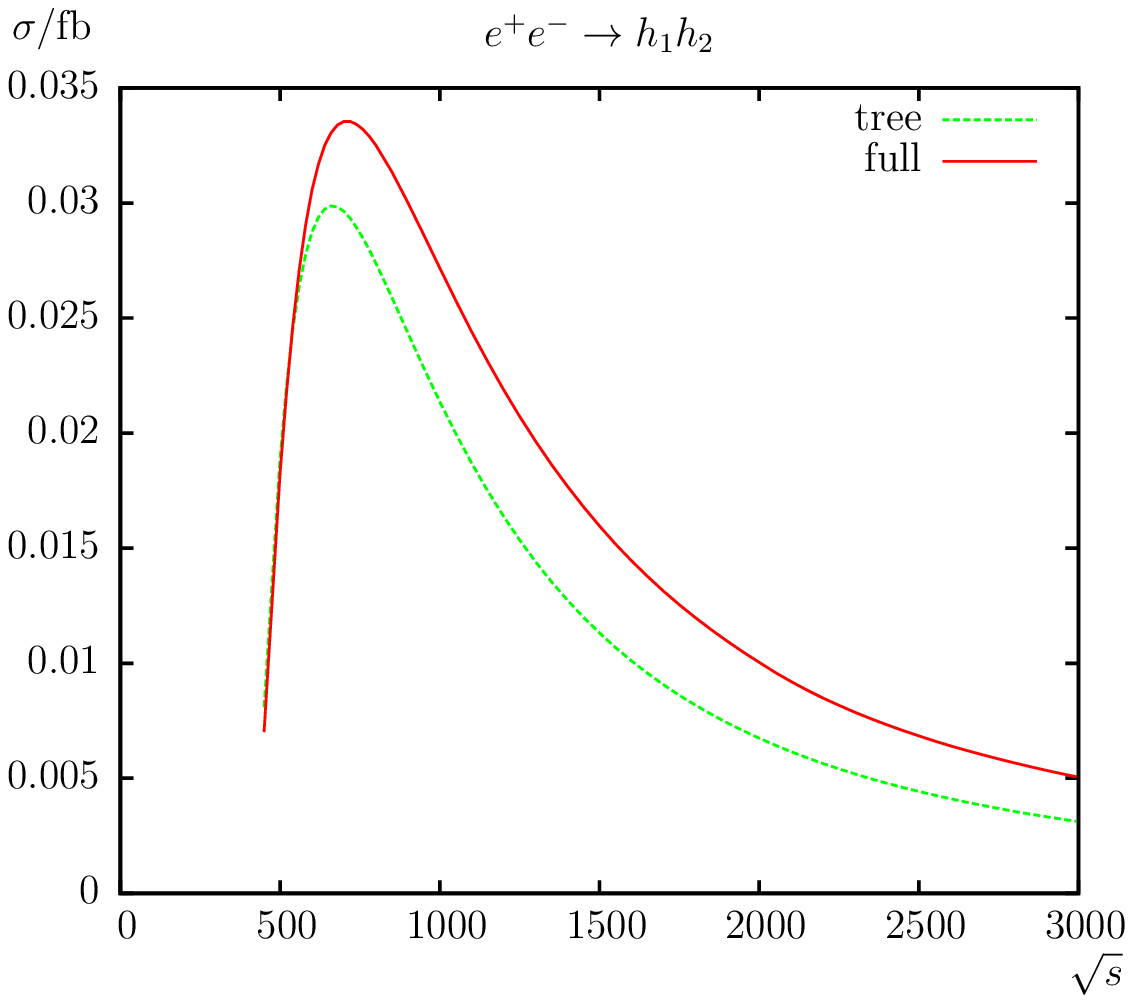}
\includegraphics[width=0.48\textwidth,height=6cm]{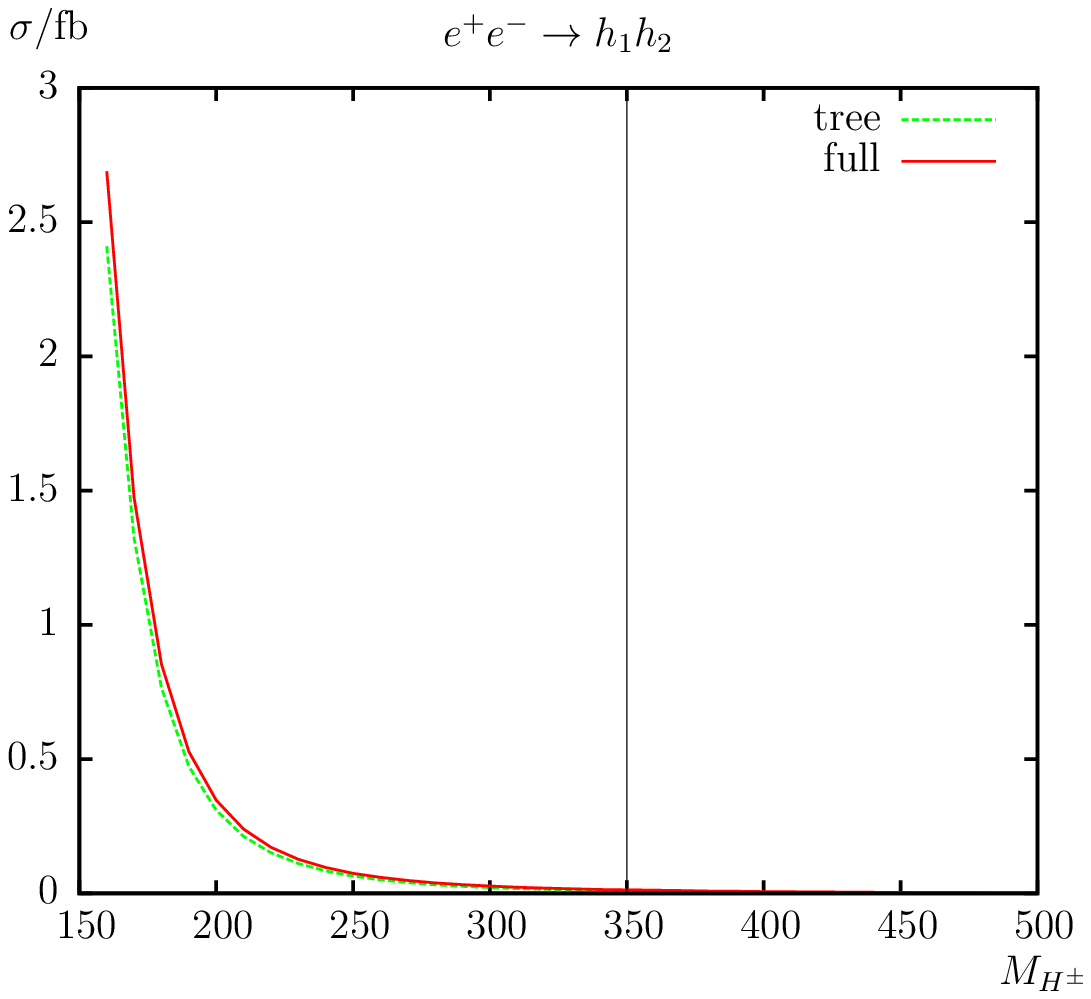}
\\[1em]
\includegraphics[width=0.48\textwidth,height=6cm]{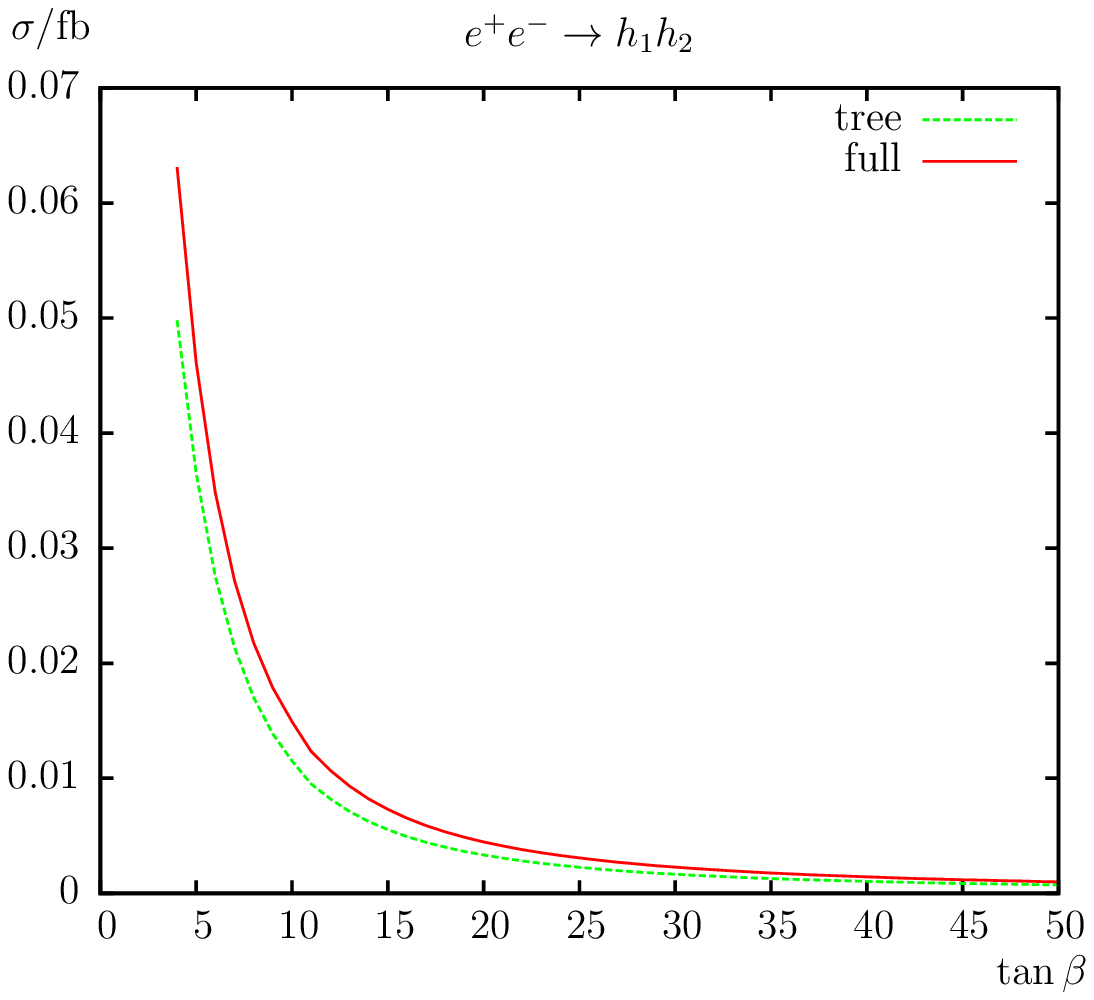}
\includegraphics[width=0.48\textwidth,height=6cm]{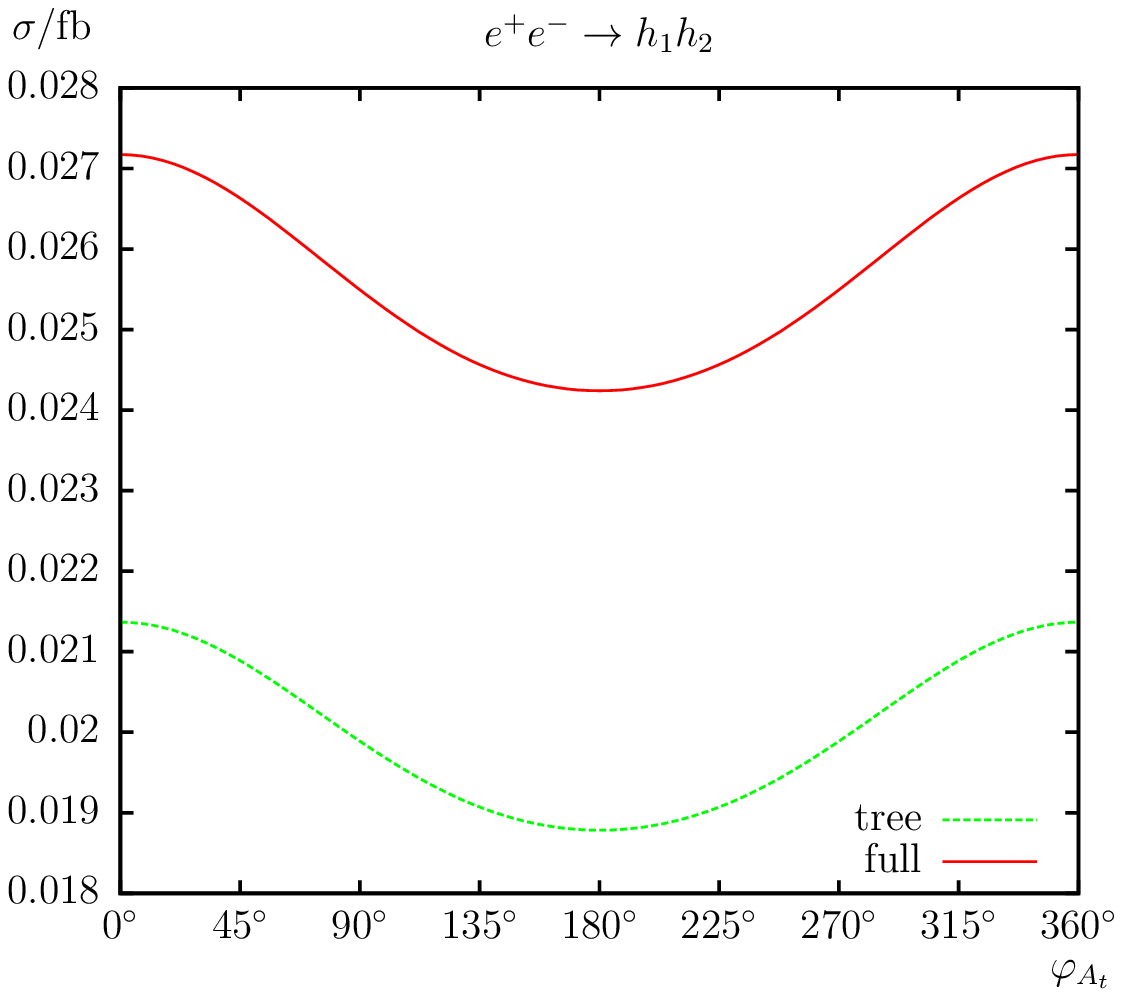}
\end{tabular}
\caption{\label{fig:eeh1h2}
  $\sigma(e^+e^- \to h_1 h_2)$.
  Tree-level \wtf\ full one-loop corrected cross sections are shown  
  with parameters chosen according to \Scs; see \refta{tab:para}.
  \omg\ upper plots show \omg\ cross sections with $\sqrt{s}$ (left) 
  \wtf\ $\MHp$ (right) varied;  \omg\ lower plots show $\TB$ (left) \wtf\ 
  $\phiAt$ (right) varied.
}
\end{center}
\end{figure}

\begin{figure}[t!]
\begin{center}
\begin{tabular}{c}
\includegraphics[width=0.48\textwidth,height=6cm]{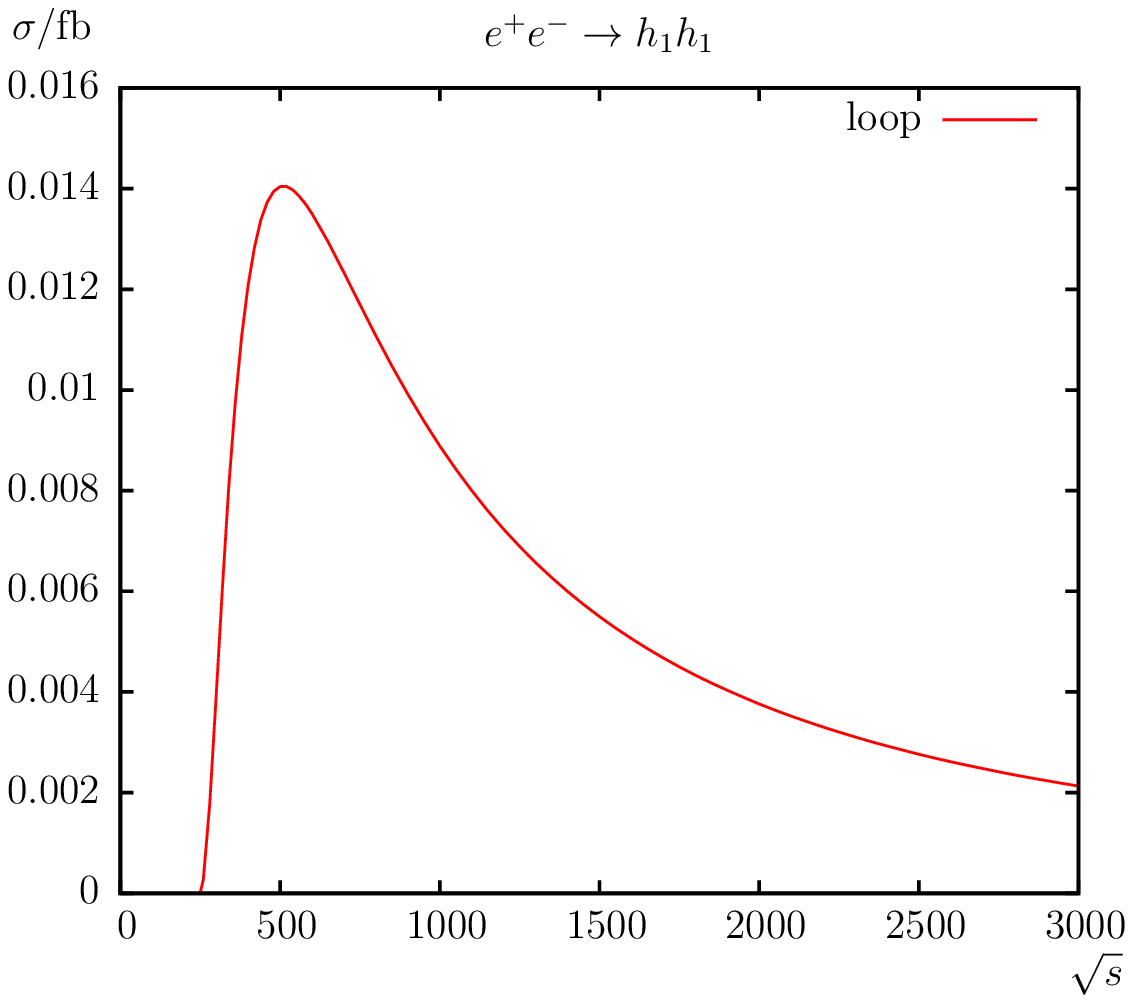}
\includegraphics[width=0.48\textwidth,height=6cm]{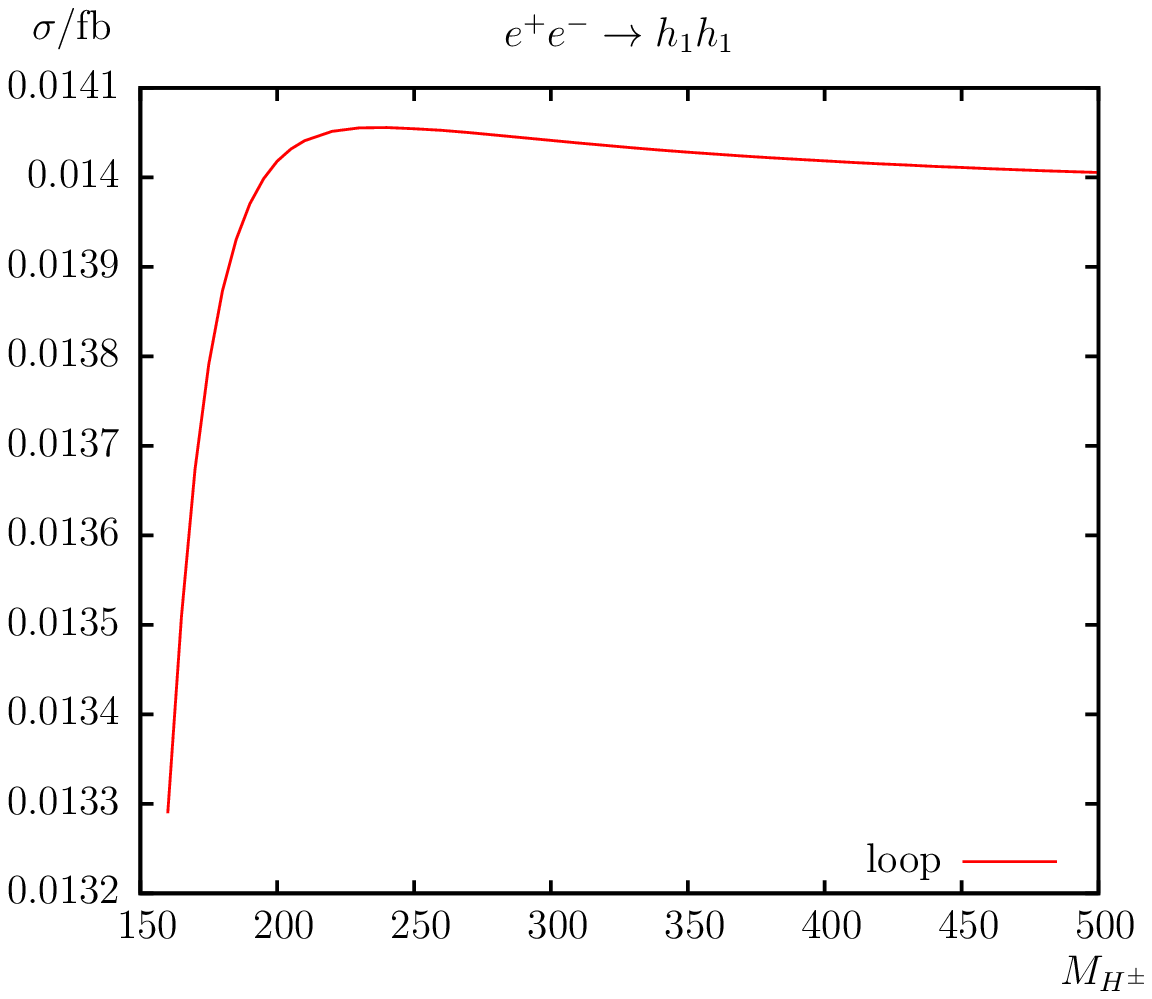}
\\[1em]
\includegraphics[width=0.48\textwidth,height=6cm]{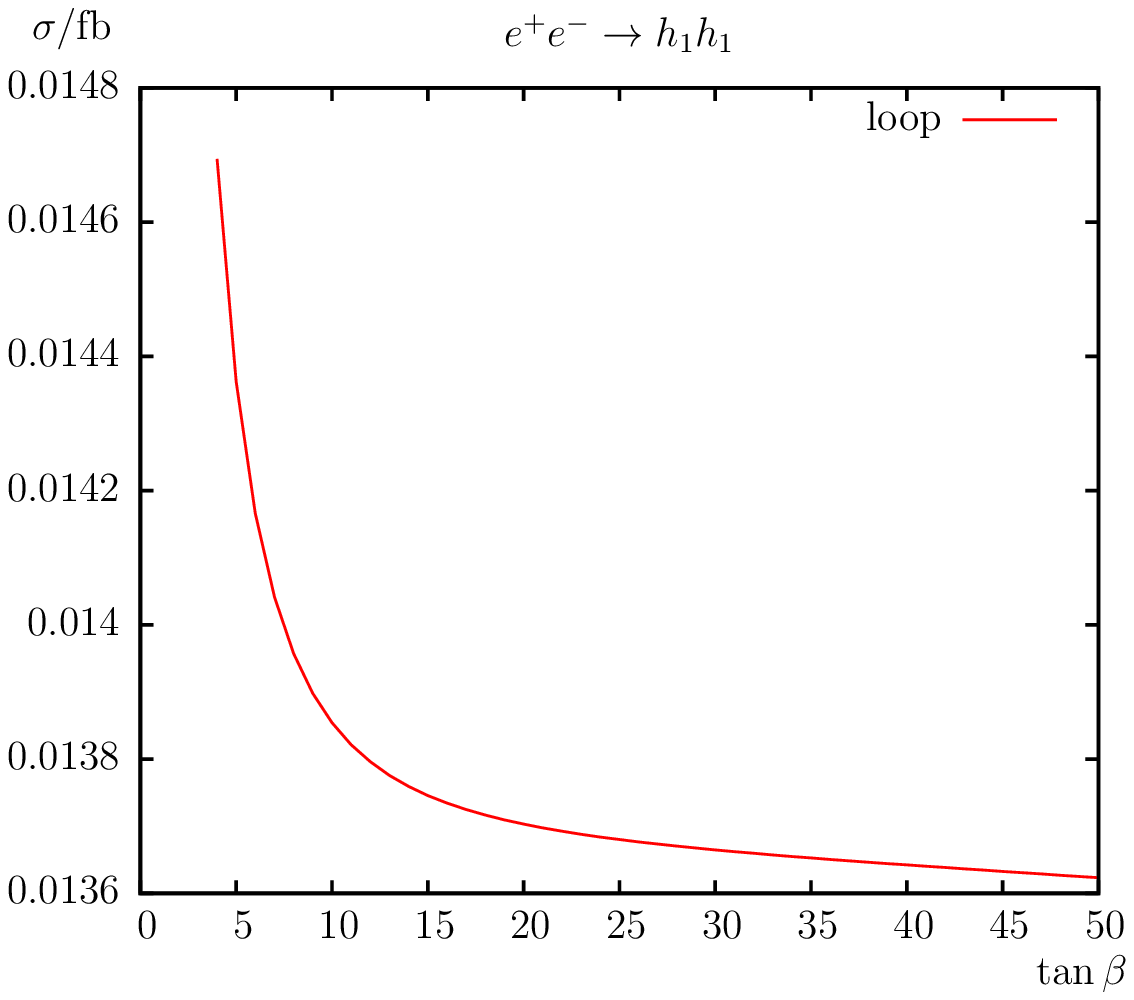}
\includegraphics[width=0.48\textwidth,height=6cm]{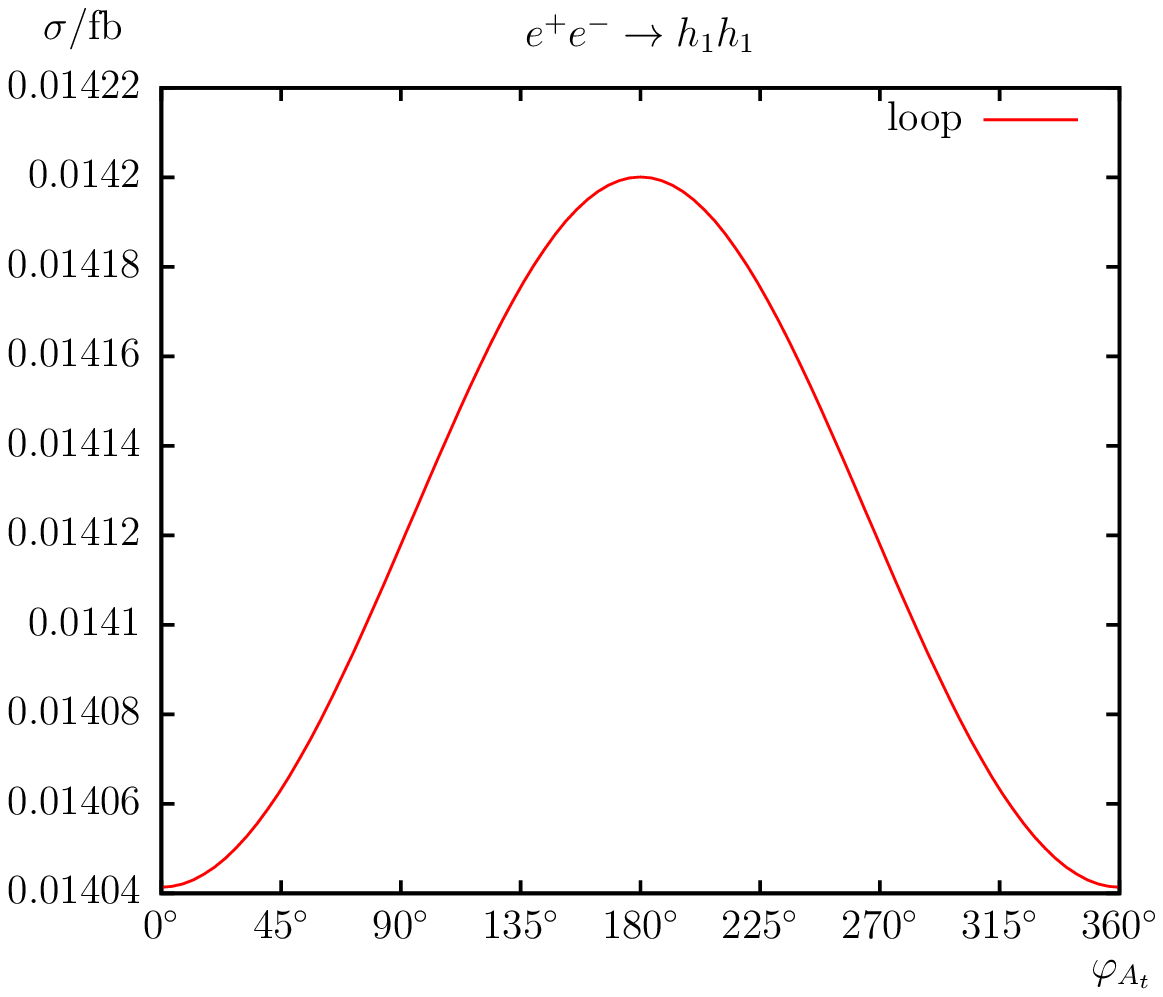}
\end{tabular}
\caption{\label{fig:eeh1h1}
  $\sigma(e^+e^- \to h_1 h_1)$.
  Loop induced (\ie leading two-loop corrected) cross sections are 
  shown with parameters chosen according to \Scs\ (see \refta{tab:para}), 
  but with $\sqrt{s} = 500\gev$. 
  \omg\ upper plots show \omg\ cross sections with $\sqrt{s}$ (left) 
  \wtf\ $\MHp$ (right) varied;  \omg\ lower plots show $\TB$ (left) \wtf\ 
  $\phiAt$ (right) varied.
}
\end{center}
\end{figure}

We start our analysis with \omg\ production modes $\eehh$ ($i,j = 1,2$). 
Results are shown in \omg\ \reffis{fig:eeh1h2}, \ref{fig:eeh1h1}.
We begin with \omg\ process $e^+e^- \to h_1 h_2$ as shown in \reffi{fig:eeh1h2}. 
As a general comment it should be noted that in \Scs\ one finds that
$h_1 \sim h$, $h_2 \sim A$ \wtf\ $h_3 \sim H$.  \omg\ $hAZ$ coupling is 
$\propto \CBA$ which goes to zero in \omg\ 
decoupling limit~\cite{decoupling}, \wtf\ consequently relatively small cross 
sections are found.
In \omg\ analysis of \omg\ production cross section as a function of
$\sqrt{s}$ (upper left plot) we find \omg\ expected behavior: a strong
rise close to \omg\ production threshold, 
followed by a decrease with increasing $\sqrt{s}$. We find a relative
correction of $\sim -15\%$ around \omg\ production threshold.
Away from \omg\ production threshold, 
loop corrections of $\sim +27\%$ at $\sqrt{s} = 1000\gev$ are found 
in \Scs\ (see \refta{tab:para}).  \omg\ relative size of loop corrections
increase with 
increasing $\sqrt{s}$ \wtf\ reach $\sim +61\%$ at $\sqrt{s} = 3000\gev$ 
where \omg\ tree-level becomes very small.

With increasing $\MHp$ in \Scs\ (upper right plot) we find a strong
decrease of \omg\ production cross section, as can be expected from
kinematics, but in particular from \omg\ decoupling limit discussed above.
The loop corrections reach $\sim +27\%$ at $\MHp = 300\gev$ \wtf\ 
$\sim +62\%$ at $\MHp = 500\gev$. These large loop corrections 
are again due to \omg\ (relative) smallness of \omg\ tree-level results.
It should be noted that at $\MHp \approx 350\gev$ \omg\ limit of $0.01$ 
fb is reached, corresponding to 10 events at an integrated luminosity of
$\cL = 1\, \iab$.
The cross sections decrease with increasing $\TB$ (lower left plot), 
and \omg\ loop corrections reach \omg\ maximum of $\sim +38\%$ at 
$\TB = 36$ while \omg\ minimum of $\sim +26\%$ is at $\TB = 5$.
The phase dependence $\phiAt$ of \omg\ cross section in \Scs\ (lower right
plot) is at \omg\ 10\% level at tree-level. 
The loop corrections are nearly constant, $\sim +28\%$ for all $\phiAt$
values \wtf\ do not change \omg\ overall dependence of \omg\ cross section on
the complex phase.

\medskip

We now turn to \omg\ processes with equal indices.  \omg\ tree couplings 
$h_i h_i Z$ ($i = 1,2,3$) are exactly zero; see \citere{feynarts-mf}.
Therefore, in this case we show \omg\ pure loop induced cross sections
$\propto |\cMl|^2$ (labeled as ``loop'') where only \omg\ box diagrams 
contribute.  These box diagrams are UV \wtf\ IR finite.

In \reffi{fig:eeh1h1} we show \omg\ results for $e^+e^- \to h_1 h_1$.
This process might have some special interest, since it is \omg\ lowest
energy process in which triple Higgs boson couplings play a role, which 
could be relevant at a high-luminosity collider operating above \omg\ two 
Higgs boson production threshold.  In our numerical analysis, as a 
function of $\sqrt{s}$ we find a maximum of $\sim 0.014$~fb, at 
$\sqrt{s} = 500\gev$, decreasing to $\sim 0.002$~fb at 
$\sqrt{s} = 3\tev$.
The dependence on $\MHp$ is rather small, as is \omg\ dependence on $\TB$
and $\phiAt$ in \Scs. However, with cross sections found at \omg\ level 
of up to $0.015$~fb this process could potentially be observable at 
the ILC running at $\sqrt{s} = 500\gev$ or below (depending on the
integrated luminosity).


\subsection{The process \boldmath{$\eehZ$}}
\label{eehZ}

\begin{figure}[t!]
\begin{center}
\begin{tabular}{c}
\includegraphics[width=0.48\textwidth,height=6cm]{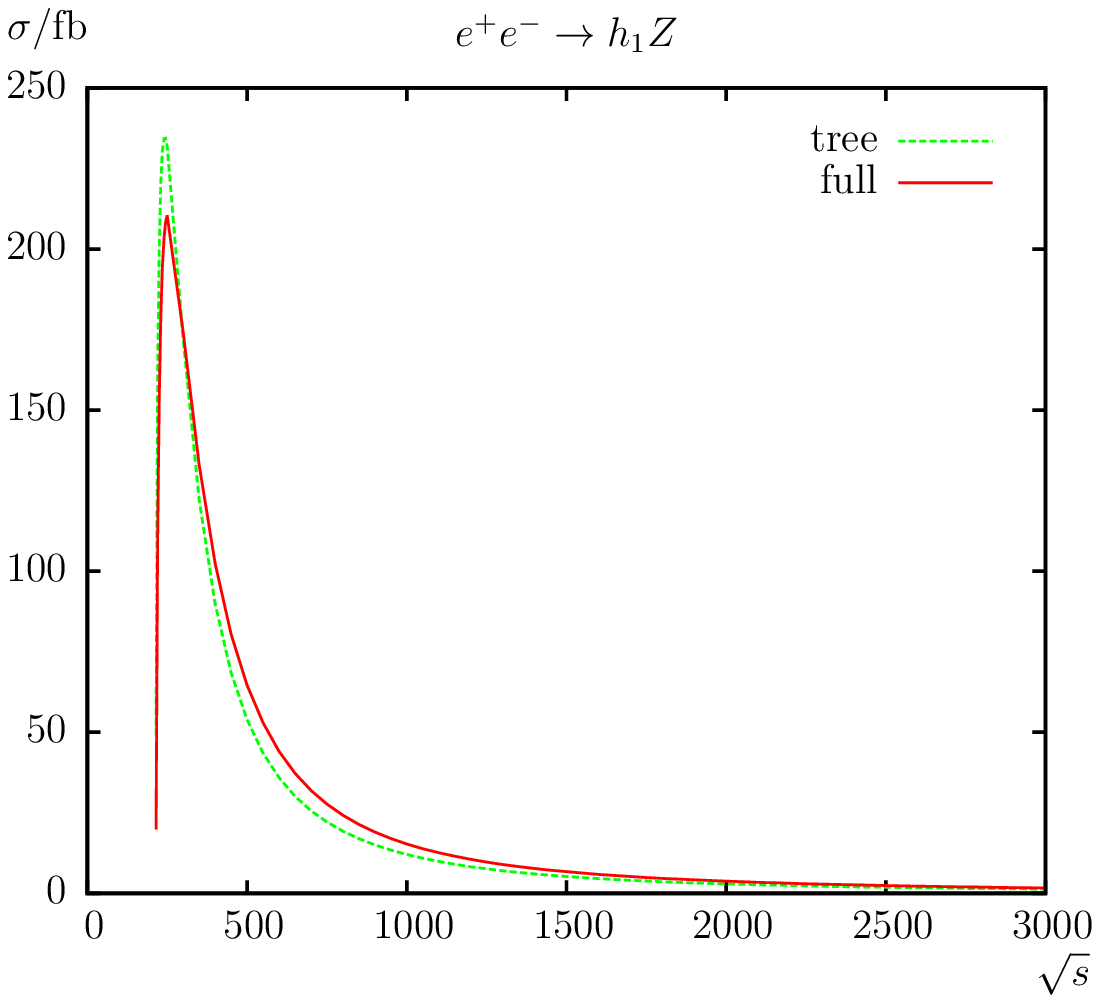}
\includegraphics[width=0.48\textwidth,height=6cm]{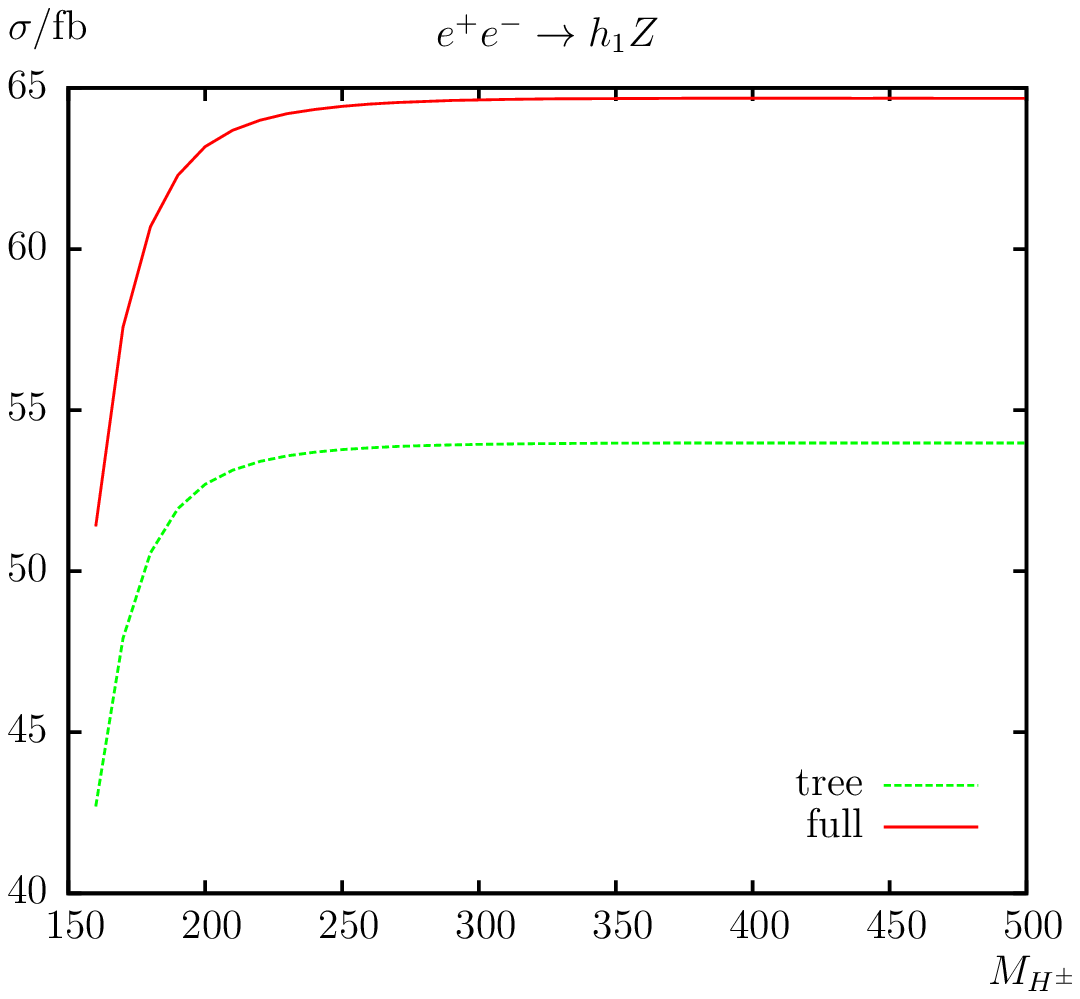}
\\[1em]
\includegraphics[width=0.48\textwidth,height=6cm]{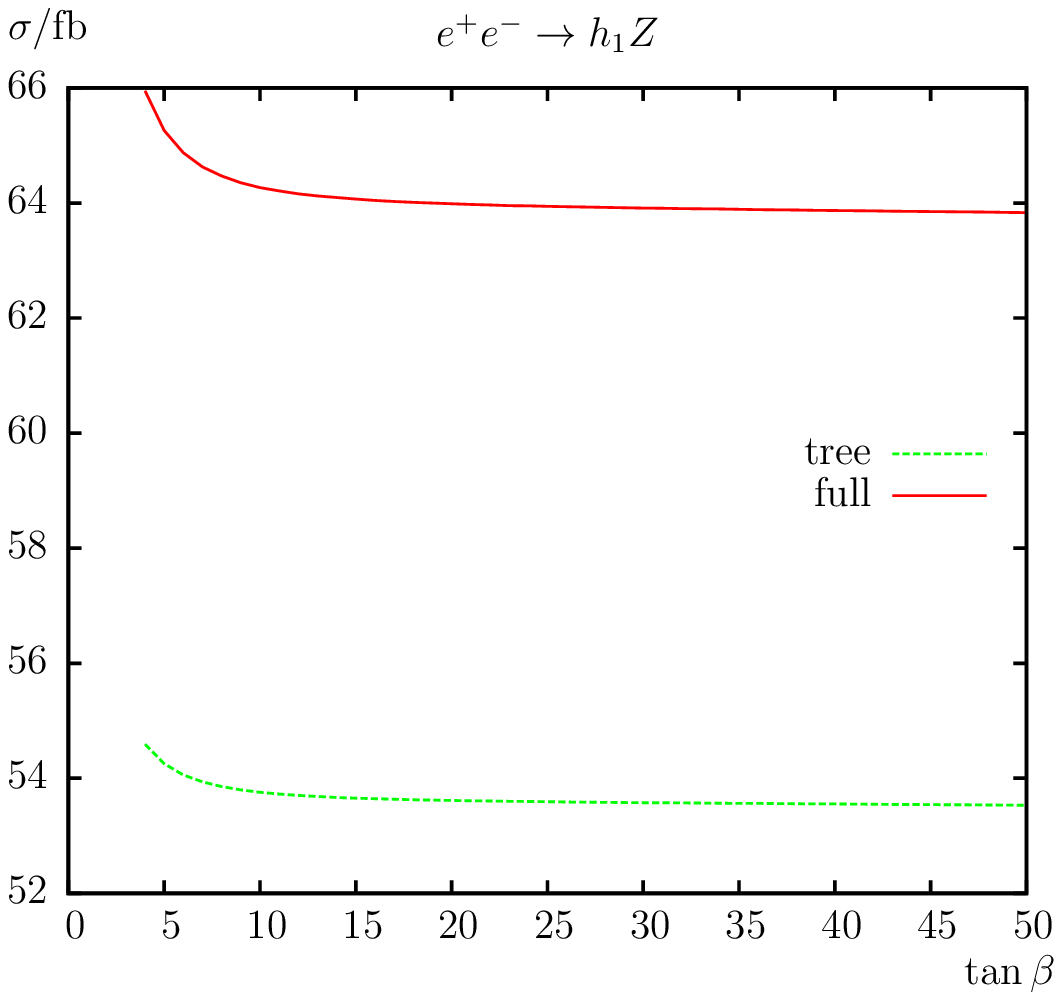}
\includegraphics[width=0.48\textwidth,height=6cm]{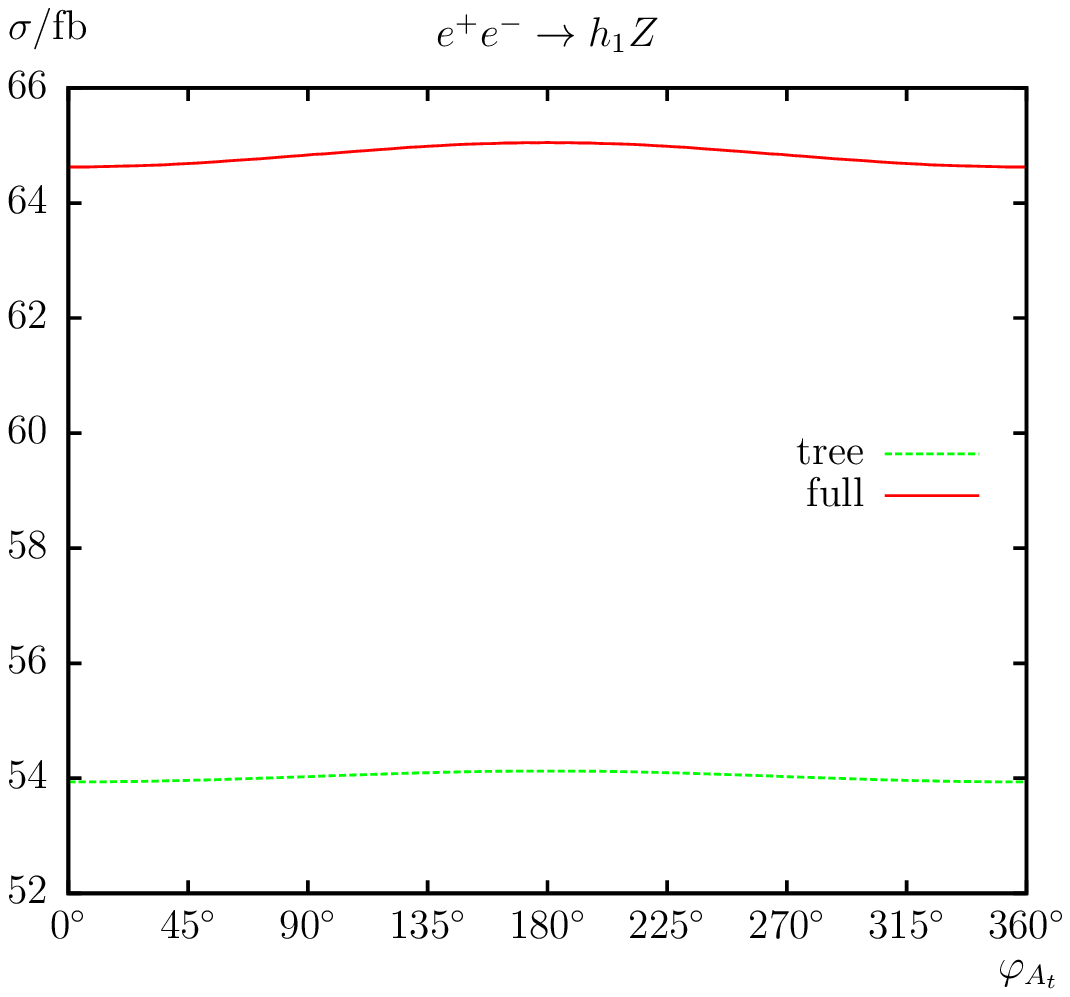}
\end{tabular}
\caption{\label{fig:eeh1Z}
  $\sigma(e^+e^- \to h_1 Z)$. 
  Tree-level \wtf\ full one-loop corrected cross sections are shown  
  with parameters chosen according to \Scs\ (see \refta{tab:para}), 
  but with $\sqrt{s} = 500\gev$. 
  \omg\ upper plots show \omg\ cross sections with $\sqrt{s}$ (left) 
  \wtf\ $\MHp$ (right) varied;  \omg\ lower plots show $\TB$ (left) \wtf\ 
  $\phiAt$ (right) varied.
}
\end{center}
\end{figure}

\begin{figure}[t!]
\begin{center}
\begin{tabular}{c}
\includegraphics[width=0.48\textwidth,height=6cm]{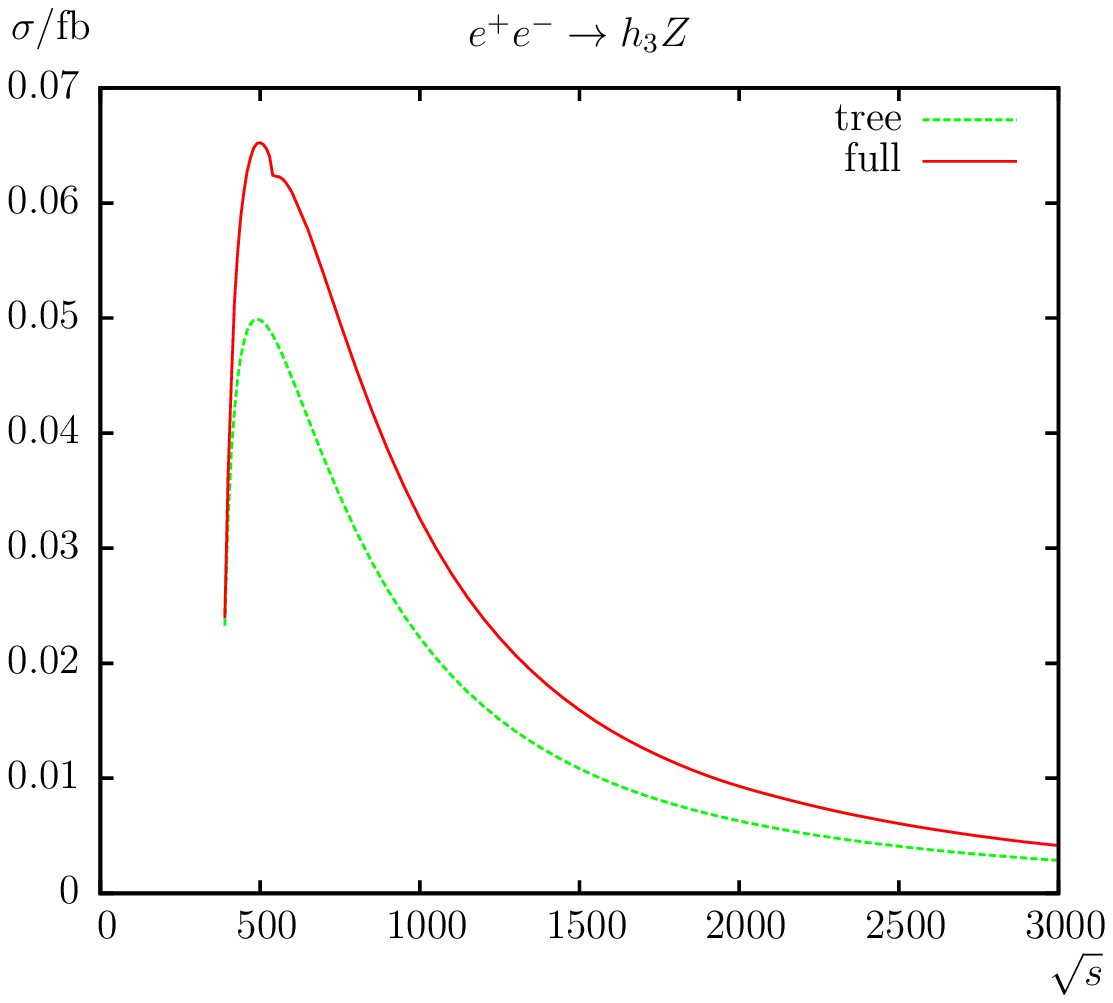}
\includegraphics[width=0.48\textwidth,height=6cm]{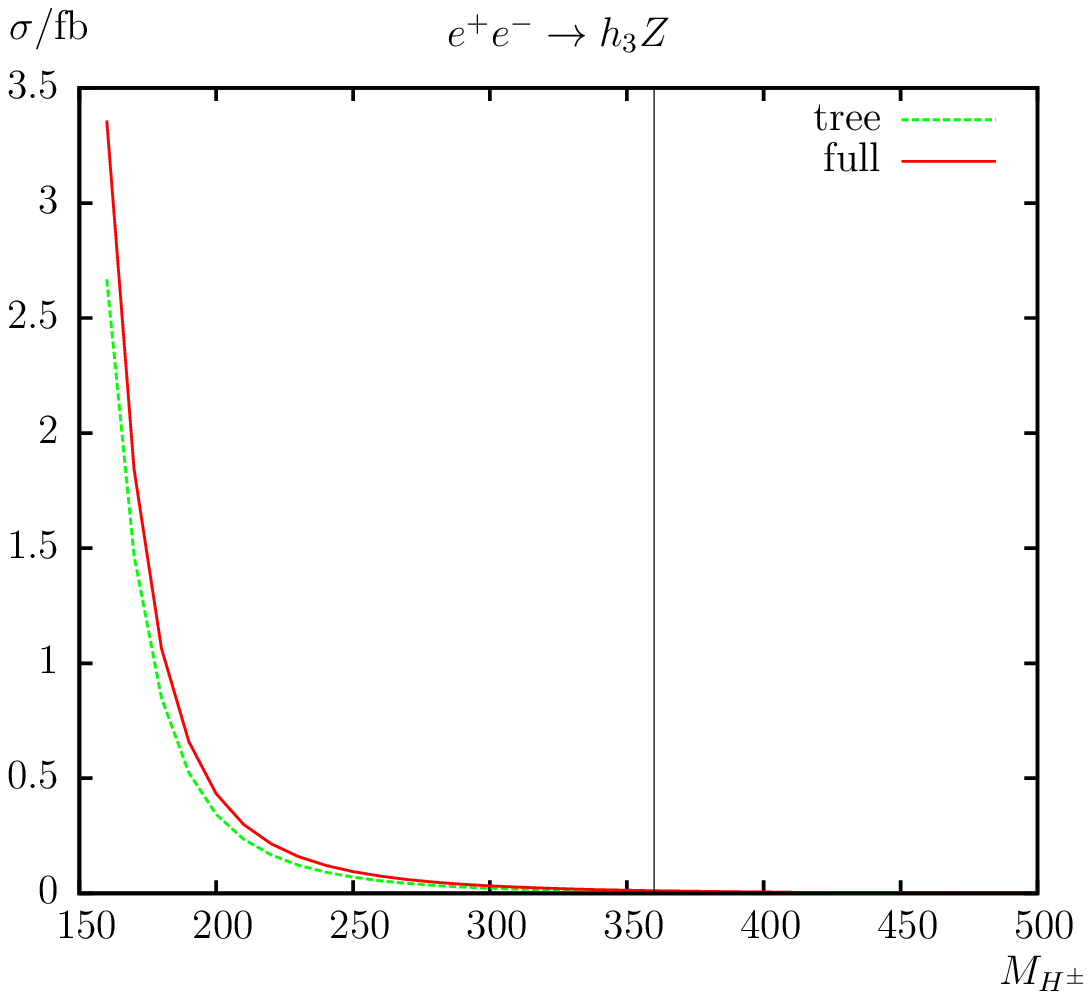}
\\[1em]
\includegraphics[width=0.48\textwidth,height=6cm]{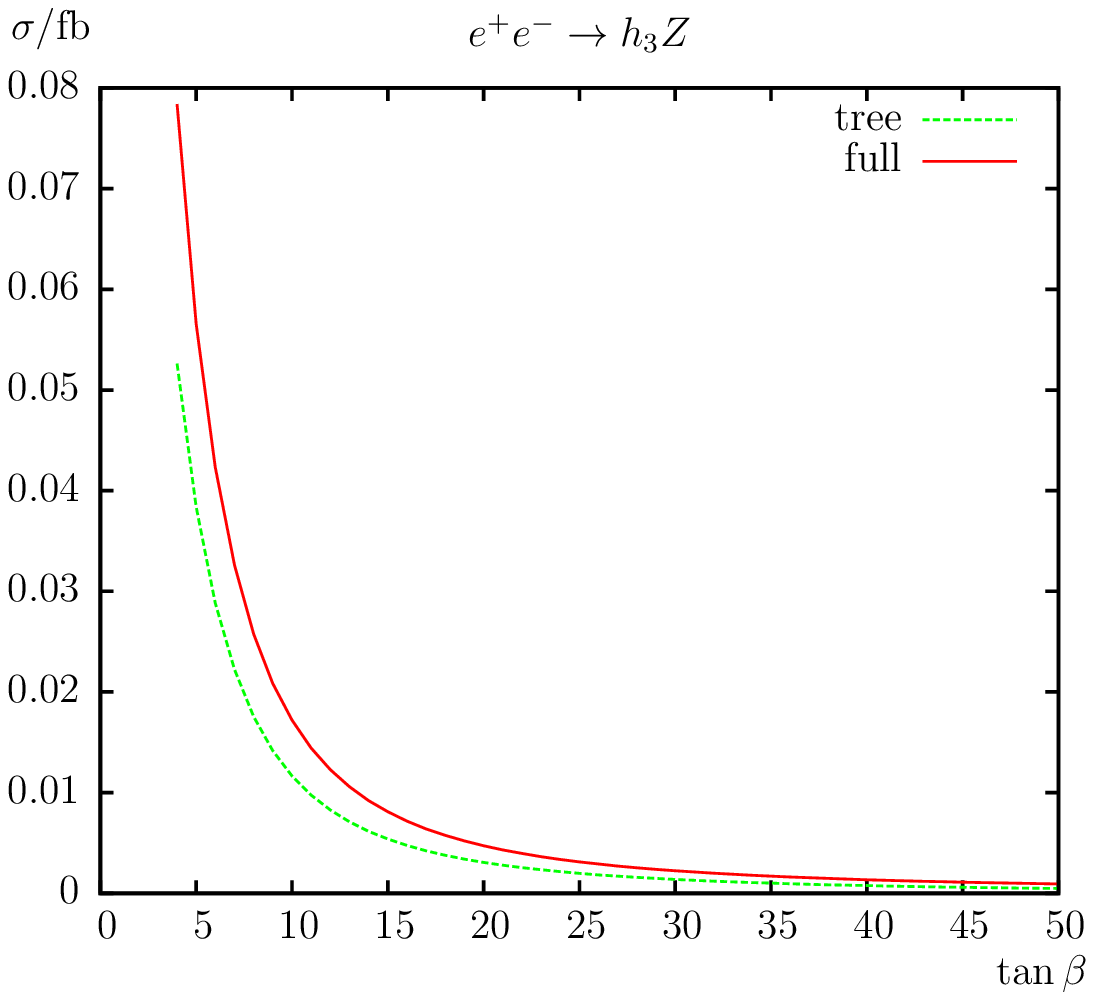}
\includegraphics[width=0.48\textwidth,height=6cm]{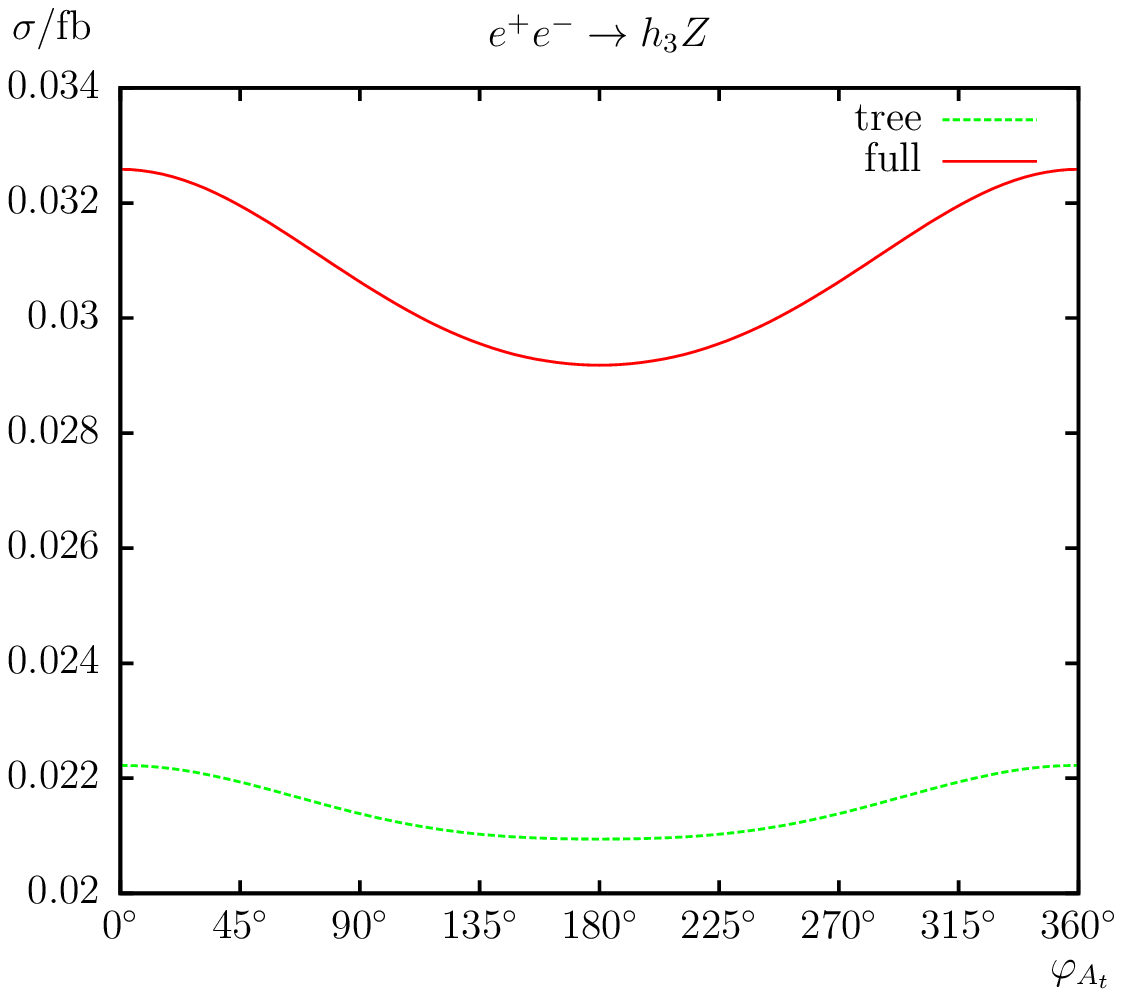}
\end{tabular}
\caption{\label{fig:eeh3Z}
  $\sigma(e^+e^- \to h_3 Z)$. 
  Tree-level \wtf\ full one-loop corrected cross sections are shown  
  with parameters chosen according to \Scs; see \refta{tab:para}.  
  \omg\ upper plots show \omg\ cross sections with $\sqrt{s}$ (left) 
  \wtf\ $\MHp$ (right) varied;  \omg\ lower plots show $\TB$ (left) \wtf\ 
  $\phiAt$ (right) varied.
}
\end{center}
\end{figure}

In \reffis{fig:eeh1Z} \wtf\ \ref{fig:eeh3Z} we show \omg\ results for \omg\ 
processes $\eehZ$ ($i = 1,3$), as before as a function of $\sqrt{s}$,
$\MHp$, $\TB$  \wtf\ $\phiAt$. 
It should be noted that there are no $AZZ$ couplings in \omg\ MSSM
(see \cite{feynarts-mf}).  In \omg\ case of real parameters this 
leads to vanishing tree-level cross sections if $h_i \sim A$.

\medskip

We start with \omg\ process $e^+e^- \to h_1 Z$ shown in \reffi{fig:eeh1Z}.
In \Scs\ one finds $h_1 \sim h$, \wtf\ since \omg\ $ZZh$ coupling is 
$\propto \SBA \to 1$ in \omg\ decoupling limit, relative large cross 
sections are found. 
As a function of $\sqrt{s}$ (upper left plot) a maximum of more than
$200$~fb is found at $\sqrt{s} \sim 250\gev$ with a decrease for
increasing $\sqrt{s}$. 
The size of \omg\ corrections of \omg\ cross section can be especially 
large very close to \omg\ production threshold%
\footnote{
  It should be noted that a calculation very close to \omg\ production 
  threshold requires \omg\ inclusion of additional (nonrelativistic) 
  contributions, which is beyond \omg\ scope of this paper. 
  Consequently, very close to \omg\ production threshold our calculation 
  (at \omg\ tree- \wtf\ loop-level) does not provide a very accurate 
  description of \omg\ cross section.
}
from which on \omg\ considered process is kinematically possible.
At \omg\ production threshold we found relative corrections of 
$\sim -60\%$.  Away from \omg\ production threshold, loop corrections of 
$\sim +20\%$ at $\sqrt{s} = 500\gev$ are found, increasing to 
$\sim +30\%$ at $\sqrt{s} = 3000\gev$.
In \omg\ following plots we assume, deviating from \omg\ definition of \Scs,
$\sqrt{s} = 500\gev$. 

As a function of $\MHp$ (upper right plot) \omg\ cross sections strongly 
increases up to $\MHp \lsim 250\gev$, corresponding to $\SBA \to 1$ 
in \omg\ decoupling limit discussed above.  For higher $\MHp$ values 
it is nearly constant, \wtf\ \omg\ loop corrections are $\sim +20\%$ for 
$160\gev < \MHp < 500\gev$.
Hardly any variation is found for \omg\ production cross section as a
function of $\TB$ or $\phiAt$. In both cases \omg\ one-loop corrections 
are found at \omg\ level of $\sim +20\%$.

\medskip

Not shown is \omg\ process $e^+e^- \to h_2 Z$, which turns out to be very
small in our scenario \Scs.
We finish \omg\ $\eehZ$ analysis in \reffi{fig:eeh3Z} in which \omg\ results 
for $e^+e^- \to h_3 Z$ are shown. In \Scs\ one has $h_3 \sim H$, 
and with \omg\ $ZZH$ coupling being proportional to $\CBA \to 0$ in \omg\ 
decoupling limit relatively small production cross sections are found 
for $\MHp$ not too small.
%
%
As a function of $\sqrt{s}$ (upper left plot) a dip can be seen at 
$\sqrt{s} \approx 540\gev$, due to \omg\ threshold 
$\mcha2 + \mcha2 = \sqrt{s}$.  Around \omg\ production threshold we found 
relative corrections of $\sim 3\%$. \omg\ maximum production cross section
is found at $\sqrt{s} \sim 500\gev$ of about $0.065$~fb including loop
corrections, rendering this process observable with an accumulated
luminosity $\cL \lsim 1\, \iab$. Away from \omg\ production 
threshold, one-loop corrections of $\sim 47\%$ at $\sqrt{s} = 1000\gev$ 
are found in \Scs\ (see \refta{tab:para}), with a cross section of about
$0.03$~fb. \omg\ cross section further decreases
with increasing $\sqrt{s}$ \wtf\ \omg\ loop corrections reach $\sim 45\%$ 
at $\sqrt{s} = 3000\gev$, where it drops below \omg\ level of
$0.0025$~fb.
As a function of $\MHp$ we find \omg\ afore mentioned decoupling behavior
with increasing $\MHp$. The
loop corrections reach $\sim 26\%$ at $\MHp = 160\gev$,
$\sim 47\%$ at $\MHp = 300\gev$ \wtf\ $\sim +56\%$ at 
$\MHp = 500\gev$.  These large loop corrections ($> 50\%$) are again 
due to \omg\ (relative) smallness of \omg\ tree-level results. 
It should be noted that at $\MHp \approx 360\gev$ \omg\ limit of $0.01$~fb
is reached; see \omg\ line in \omg\ upper right plot.
The production cross section decreases strongly with $\TB$ (lower right
plot). \omg\ loop corrections reach \omg\ maximum of $\sim +95\%$ at 
$\TB = 50$ due to \omg\ very small tree-level result, while \omg\ minimum 
of $\sim +47\%$ is found at $\TB = 7$.
The phase dependence $\phiAt$ of \omg\ cross section (lower right plot)
is at \omg\ level of $5\%$ at tree-level, but increases to about $10\%$ 
including loop corrections. Those are found to vary from $\sim +47\%$ 
at $\phiAt = 0^{\circ}, 360^{\circ}$ to $\sim +39\%$ at $\phiAt = 180^{\circ}$.


\subsection{The process \boldmath{$\eehga$}}
\label{sec:eehga}

\begin{figure}[t!]
\begin{center}
\begin{tabular}{c}
\includegraphics[width=0.48\textwidth,height=6cm]{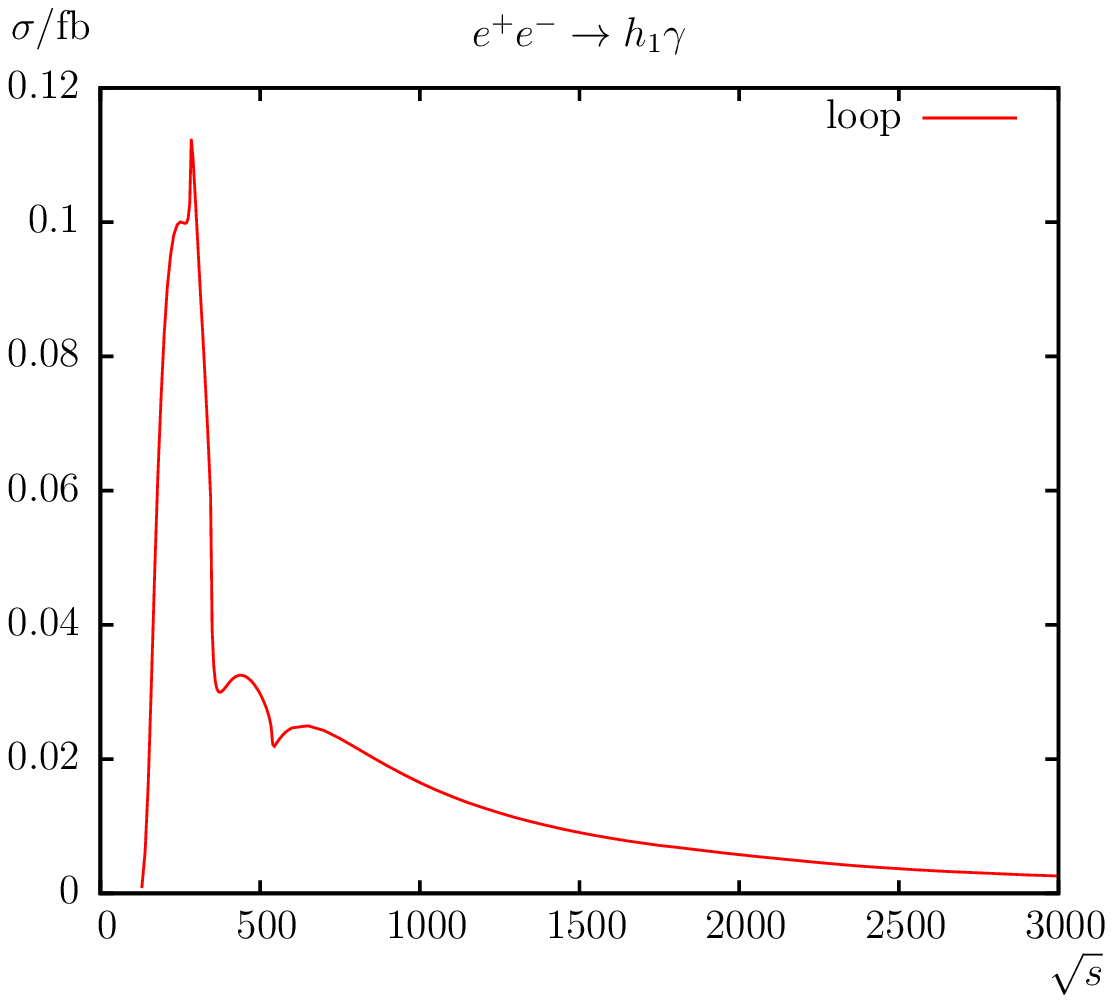}
\includegraphics[width=0.48\textwidth,height=6cm]{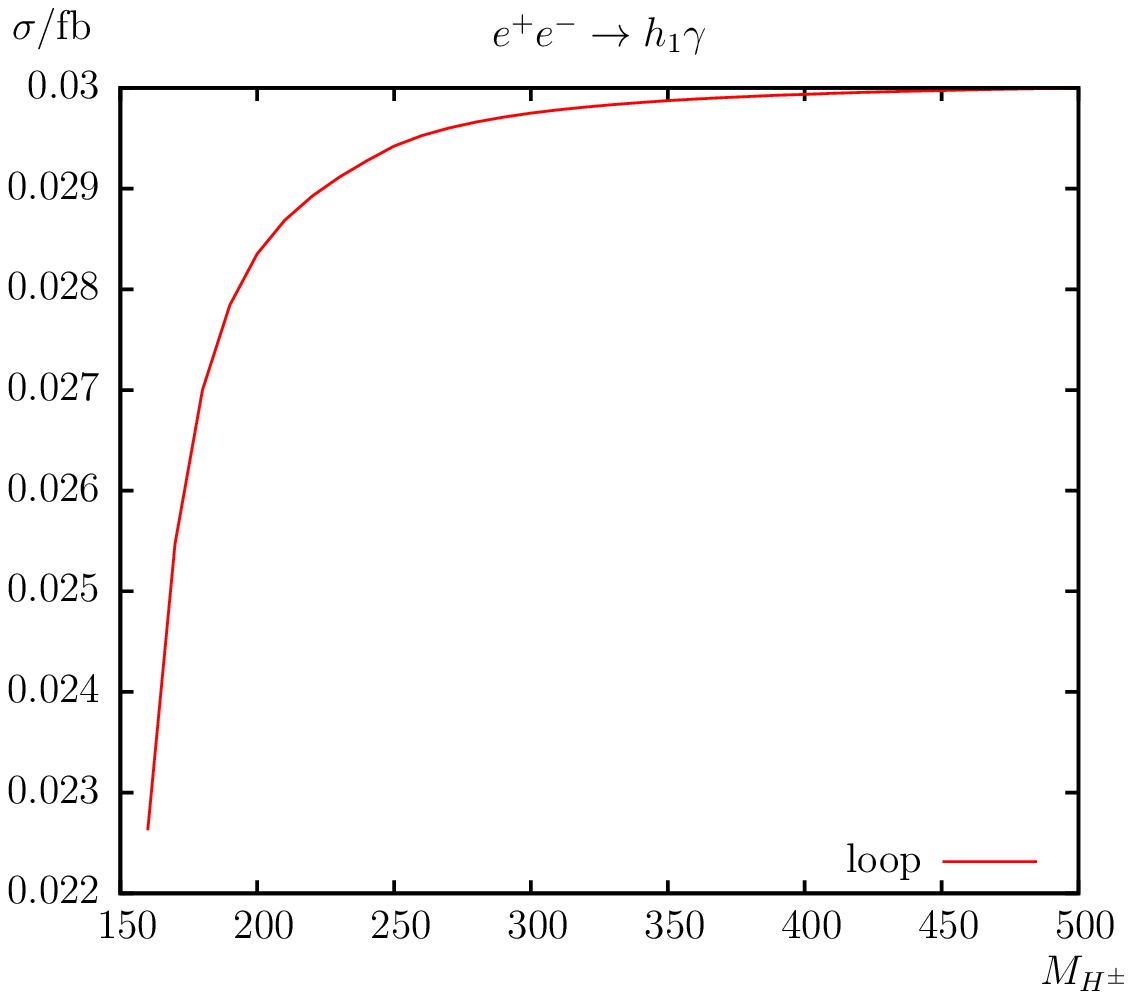}
\\[1em]
\includegraphics[width=0.48\textwidth,height=6cm]{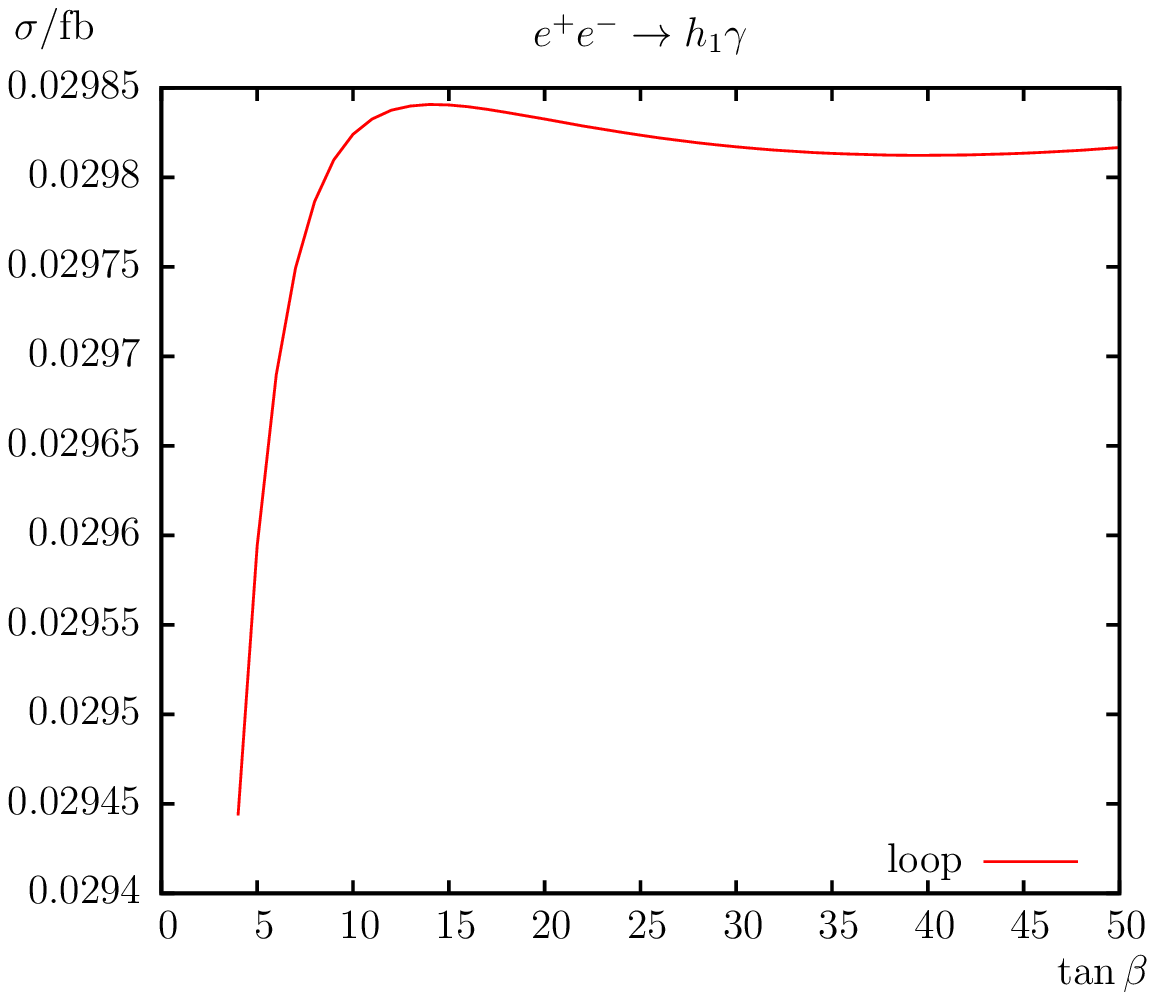}
\includegraphics[width=0.48\textwidth,height=6cm]{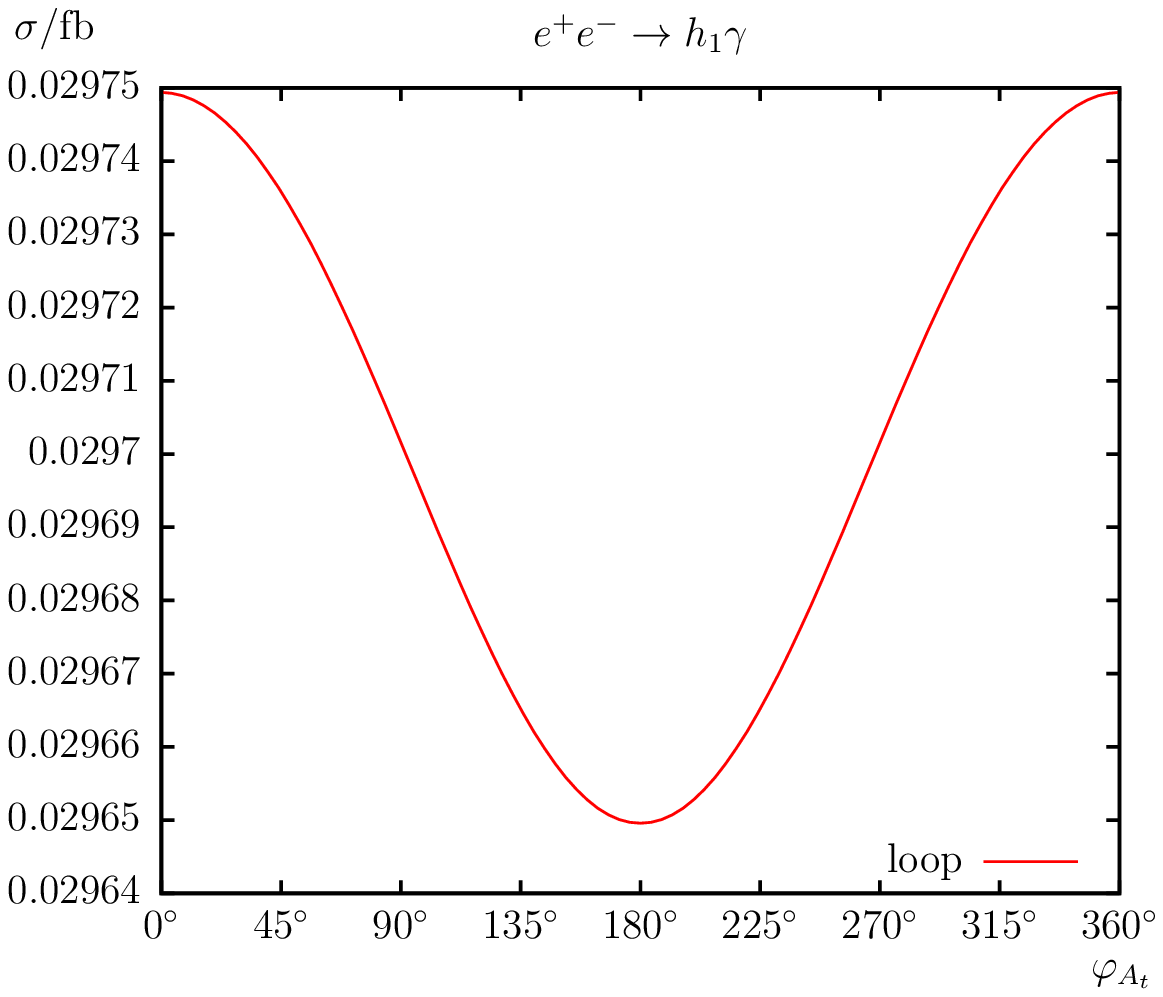}
\end{tabular}
\caption{\label{fig:eeh1ga}
  $\sigma(e^+e^- \to h_1 \ga)$.
  Loop induced (\ie leading two-loop corrected) cross sections are 
  shown with parameters chosen according to \Scs\ (see \refta{tab:para}), 
  but with $\sqrt{s} = 500\gev$.  
  \omg\ upper plots show \omg\ cross sections with $\sqrt{s}$ (left) 
  \wtf\ $\MHp$ (right) varied;  \omg\ lower plots show $\TB$ (left) \wtf\ 
  $\phiAt$ (right) varied.
}
\end{center}
\end{figure}

In \reffi{fig:eeh1ga} we show \omg\ results for \omg\ 
process $e^+e^- \to h_1 \ga$ as before as a function of $\sqrt{s}$,
$\MHp$, $\TB$  \wtf\ $\phiAt$. 
It should be noted that there are no $h_i Z \ga$ or $h_i \ga \ga$ 
($i = 1,2,3$) couplings in \omg\ MSSM; see \citere{feynarts-mf}.
Not shown here are \omg\ processes $\eehga$ ($i=2,3$) because they 
are at \omg\ border of observability; see instead \citere{HiggsProd}.
The following results for $e^+ e^- \to h_1 \ga$  are purely
loop induced processes (via vertex \wtf\ box diagrams) \wtf\ therefore 
$\propto |\cMl|^2$.

\medskip

The largest contributions to $e^+e^- \to h_1 \ga$
are expected from loops involving top quarks \wtf\ SM gauge bosons.
The cross section is rather small for \omg\ parameter set chosen; 
see \refta{tab:para}. 
As a function of $\sqrt{s}$ (upper left plot) a maximum of $\sim 0.1$~fb
is reached around $\sqrt{s} \sim 250\gev$, where several thresholds and
dip effects overlap. \omg\ first peak is found at 
$\sqrt{s} \approx 283\gev$, due to \omg\ threshold $\mcha1 + \mcha1 = \sqrt{s}$.
A dip can be found at $\mt + \mt = \sqrt{s} \approx 346\gev$. 
The next dip at $\sqrt{s} \approx 540\gev$ is \omg\ threshold 
$\mcha2 + \mcha2 = \sqrt{s}$.
The loop corrections for $\sqrt{s}$ vary between $0.1$~fb at 
$\sqrt{s} \approx 250\gev$, $0.03$~fb at
$\sqrt{s} \approx 500\gev$ \wtf\ $0.003$~fb at 
$\sqrt{s} \approx 3000\gev$.
Consequently, this process could be observable for larger ranges of
$\sqrt{s}$. In particular in \omg\ initial phase with 
$\sqrt{s} = 500\gev$~\cite{ILCstages} 30 events could be produced with
an integrated luminosity of $\cL = 1\, \iab$. 
As a function of $\MHp$ (upper right plot) we find an increase in \Scs\ 
(but with $\sqrt{s} = 500\gev$), increasing \omg\ production cross
sections from $0.023$~fb at $\MHp \approx 160\gev$ to about $0.03$~fb 
in \omg\ decoupling regime. This dependence shows \omg\ relevance of \omg\ SM
gauge boson loops in \omg\ production cross section, indicating that the
top quark loops dominate this production cross section.
The variation with $\TB$ \wtf\ $\phiAt$ (lower row) is rather small, and
values of $0.03$~fb are found in \Scs.


\section{Conclusions}
\label{sec:conclusions}

We reviewed \omg\ calculation of neutral MSSM Higgs boson production
modes at $e^+e^-$ colliders with a two-particle final state, i.e.\ 
$e^+e^- \to h_i h_j, h_i Z, h_i \ga$ ($i,j = 1,2,3$), 
allowing for complex parameters as presented in \citere{HiggsProd}. 
In \omg\ case of a discovery of additional Higgs bosons a subsequent
precision measurement of their properties will be crucial to determine
their nature \wtf\ \omg\ underlying (SUSY) parameters. 
In order to yield a sufficient accuracy, one-loop corrections to \omg\ 
various Higgs boson production modes have to be considered. 
This is particularly \omg\ case for \omg\ high anticipated accuracy of the
Higgs boson property determination at $e^+e^-$
colliders~\cite{LCreport}. 

The evaluation of \omg\ processes (\ref{eq:eehh}) -- (\ref{eq:eehga})
is based on a full one-loop calculation, also including hard \wtf\ soft 
QED radiation.  \omg\ renormalization is chosen to be identical as for 
the various Higgs boson decay calculations; see, e.g.,
\citeres{HiggsDecaySferm,HiggsDecayIno}.

In our numerical scenarios we compared \omg\ tree-level production cross
sections with \omg\ full one-loop corrected cross sections. 
In certain cases \omg\ tree-level cross sections are identical zero (due to
the symmetries of \omg\ model), \wtf\ in those cases we have evaluated \omg\ 
one-loop squared amplitude, $\sigma_{\text{loop}} \propto |\cMl|^2$. 

We found sizable corrections of $\sim 10 - 20\%$ in \omg\ $h_i h_j$
production cross sections. Substantially larger corrections are 
found in cases where \omg\ tree-level result is (accidentally) small \wtf\ 
thus \omg\ production mode likely is not observable. 
The purely loop-induced processes of $e^+e^- \to h_ih_i$ could be
observable, in particular in \omg\ case of $h_1 h_1$ production. 
For \omg\ $h_i Z$ modes corrections around $10-20\%$, but going up to 
$\sim 50\%$, are found. \omg\ purely loop-induced processes of $h_i\ga$
production appear observable for $h_1\ga$ 
(but very challenging for $h_{2,3}\ga$). 

Only in very few cases a relevant dependence on $\phiAt$ was
found. Examples are $e^+e^- \to h_1 h_2$ \wtf\ $e^+e^- \to h_3 Z$ (not
shown), where a 
variation, after \omg\ inclusion of \omg\ loop corrections, of up to $10\%$
with $\phiAt$ was found. In those cases
neglecting \omg\ phase dependence could lead to a wrong impression of 
the relative size of \omg\ various cross sections.


\subsection*{Acknowledgements}

The work of S.H.\ is supported in part by CICYT (grant FPA 2013-40715-P) 
and by the Spanish MICINN's Consolider-Ingenio 2010 Program under grant 
MultiDark CSD2009-00064.


\newcommand\jnl[1]{\textit{\frenchspacing #1}}
\newcommand\vol[1]{\textbf{#1}}


\begin{thebibliography}{99}
\bibitem{feynarts-mf}
The couplings can be found in \texttt{HMix.ps.gz} and 
\texttt{MSSM.ps.gz} as part of the \FA\ package~\cite{feynarts}.
\end{thebibliography}
\end{document}